\documentstyle[12pt,aasms4]{article}
\begin{document}

\title{Calibration and performance of the photon-counting detectors for the
Ultraviolet Imaging Telescopes (UVIT) of the Astrosat
observatory\footnote{Astrosat is an observatory project of the Indian
Space Research Organisation (ISRO), with partnership by the Canadian
Space Agency (CSA)}}

\author{J. Postma}
\affil{Dept of Physics and Astronomy, University of Calgary, Calgary,
AB, Canada}

\author{J.B. Hutchings}
\affil{Herzberg Institute of Astrophysics, 5071 West Saanich Rd.,
Victoria, B.C. V9E 2E7, Canada; john.hutchings@nrc.ca}

\author{D. Leahy}
\affil{Dept of Physics and Astronomy, University of Calgary, Calgary,
AB, Canada}

\begin{abstract}
We describe calibration data, and discuss performance of the photon-counting 
flight detectors for the Ultraviolet Imaging Telescopes on the Astrosat
observatory. The paper describes dark current, flat field and 
light-spot images for FUV, NUV, and Visible band detectors at 
more than one wavelength setting for each. We also report on nominal gain
and low-gain operations, full- and sub-window read rates, and non-photon-counting modes of operation, all expected to be used in flight. 
We derive
corrections to the event centroids from the CMOS readout arrays, for different
centroid algorihtms. We derive spatial resolution values for each
detector and plots of point-source signal saturation for different 
flux levels. We also discuss ways to correct for saturation in extended 
object images. 
\end{abstract}

\keywords{Instrumentation: detectors --- Space vehicles: instruments --- Ultraviolet:general}

\section{Introduction and data}

Astrosat is a multi-wavelength space observatory of the Indian Space Research
Organisation (ISRO). The satellite is to be launched in 2012, and contains 
three pointed X-ray instruments and two UV-optical telescopes, all with fields 
of view that are aligned. There is also an X-ray scanning sky monitor. 
The full observatory details and capability are described in the ISRO 
web pages. 

The observatory constitutes a multi-wavelength capability that allows
simultaneous monitoring of targets from 100 Kev to optical wavelengths,
with high timing precision. It is intended as a facility with proposal
time for all, as well as guaranteed time for the instrument teams. Data
from all instruments will be available for all observations. The sky
monitor will also be a trigger for target of opportunity observations. 
The mission lifetime is 5 years, which will be extended as
long as it remains in good operational order. 

   The twin UV-optical telescopes (UVIT)  have 38cm aperture and field of 
view of 29 arcmin with roughly 1 arcsec spatial resolution. One telescope 
has FUV capability and the other covers NUV and blue-optical by means
of a beamsplitter. Thus, three wavelength channels are observed 
simultaneously.  Each channel has a filter wheel that contains interference 
or longpass filters. The UV channnels also have gratings, and the optical 
has a neutral density filter. 

   The detector details and their basic operational modes are described 
in Hutchings et al (2009: paper 1), and will not be repeated here. Briefly,
the detectors are photon-counting devices, with different photocathodes
for each wavelength channel, but otherwise identical. The photcathodes
- FUV (CsI), NUV (CsTe), and VIS (S20) -
are on the inside of 40mm MgF$_2$2 windows, releasing electrons to be
accelerated across a small gap to a microchannel plate stack. The resulting electron cloud illuminates a phosphor, whose light passes through
a fibreglass taper, to be read out by a CMOS chip with 512 pixels on a side.
Photon events are centroided to 1/8 pixel to accumulate the final science
images. The systems can also be run in `integrate' mode when photon
fluxes are too high - principally in the optical channel - which lowers the
spatial resolution, but allows centroding to sufficient precision
to track and compensate for spacecraft drift, in accumulating
the final UV images. There are three different centroiding algorithms
to choose from, and a bright-object shutdown. The full field is
read at 29 Hz, but subarrays can be read up to 600 Hz for bright
targets or for finer time resolution.  

These detectors differ from those flown on FUSE and GALEX in having 
a readout array of pixels, rather than delay lines (e.g. Morrissey et al
2009, 2011, Sahnow et al 2000). They also differ from
the MAMA and delay-line readouts of the STIS and COS detectors 
(Vallerga et al 2002, 
McPhate et al 2010). Pixel-based readouts introduce systematic effects 
in the centroiding and counting of photon events, which require some 
calibration and testing (paper 1, and Kuin and Rosen 2008, Breeveld et al
2010).

 One engineering unit (NUV) and one flight unit for each channel were
fabricated by Routes Astro-engineering. Functional tests were performed 
at the Canadian Space Agency David Florida Lab (DFL), and then were fully 
tested and calibrated in a specially equipped UV vacuum facility at the
University of Calgary. This paper describes the calibration data and 
their analysis. The paper should be of interest in describing the 
functionality of the detectors and also for those interested in making
observations with the instrument in orbit.

We reported in paper 1 on results from a laboratory version of the
photon-counting detector system for the UVIT telescopes. 
In this paper we describe the calibration and performance of the 
engineering and flight hardware. The results we describe show the
performance of the systems in various modes of operation, and should
be useful in designing observing programs with UVIT.  

The FUV system has its own flight telescope, and the NUV and VIS share the
focal plane of a second telescope, via a beamsplitter. The 38cm telescopes
are designed to deliver images of 1" resolution, but the VIS images
will be degraded somewhat by passing through the beamsplitter. Each detector
will have filters or gratings on a wheel in the converging beam. All channels will operate simultaneously and have the same (29') field of view. The UV 
channels are expected to operate mainly as photon-counters, although they all
are capable of `integrate' observations by summing the raw CMOS readout 
frames. The visible channel is intended mainly for monitoring spacecraft
drift by centroiding on a star image using integrate mode, but may also
be useful for multi-wavelength observations of variable objects, when
used in fast-read photon-counting mode. 

The system functional description was given in paper 1. Individual 
`front ends' - the windows, MCP, taper, and CMOS readout - were assembled
after selection of individual components for engineering, flight, and spares
units. Considerations in the selections were the window-MCP gap, which
affects spatial resolution, photocathode uniformity and QE, and the
cosmetic cleanliness of the optical fibre tapers.  The engineering unit
had a NUV photocathode and larger than optimal window gap. Each front-end
assembly had data from a range of operating gap and MCP gain (voltages) to
enable the choice of the optimum values for each. In addition, a number of
spatial resolution tests were done on prototype front ends, to characterize 
the dependence on window voltage and wavelength.

The DFL tests yielded dark frames and pulse-height distributions
for detectors at their nominal voltage settings. The event centroiding
algorithms built into the flight units are the 3-cross, 3-square, and 
5-square subarrays from those described in paper 1. A settable bright-object
protection threshold is also built in, which shuts down the high voltage, 
if it is exceeded in adjacent events in the readout. 

The Calgary lab is built about a small, dark vacuum tank, with ports 
that allow light from lamps with a monochromator. The light flux can be
controlled by an aperture, and the level is monitored by a UV-sensitive
photometer through a beamsplitter from the main beam. The light can
be collimated, but the collimated beam is smaller than the detector windows,
and not very uniform. Uniform illumination for flat fields was available
using an integrating sphere within the tank, fed by an uncollimated beam.

Molecular contamination is important to control for UV-sensitive detectors,
and the entire lab was kept clean, and entered via an air-shower. Hardware
was handled on a clean flow bench. The detectors were
mounted in the tank, along with their flight power supplies, on a platform
that moves in 3 axes, and were operated by ground support equipment (GSE)
outside the tank. Data were viewed and logged via the GSE, and final 
images were derived later in a separate computer system.

We describe below the different test results and their implications for
science use on Astrosat.

\section{Darks and read rates}

In photon-counting operation, the dark current behaviour is a peripheral
concern. But for the UVIT detector system, a quasi-integrate mode feature was 
also developed, in which it is very important to understand the behaviour 
of the readout dark current  At nominal gain, a single detected photon 
event will utilize the maximum amount of CMOS ADC bit depth without 
saturating the detector in order to maximize counting and centroiding 
fidelity.  Therefore two photons detected in one place within a single
read-frame will saturate the CMOS.  This is analogous to pulse pile-up 
on single channel photometers.  At normal high-voltage operating 
parameters, the frames are read at approximately 30Hz for full-frame 
CMOS scanning (512 x 512 pixels), and up to approximately 600Hz for a 
sub-scan of 100 x 100 CMOS pixels. Increasing 
the read rate to 600Hz from 30Hz will allow brighter objects 
to be imaged, extending the dynamic range by more than 3 magnitudes.

 At all these rates, the background in a single 
CMOS scan is basically a bias frame, as there is too little time to allow 
any appreciable build-up of dark current.  The local background bias 
surrounding a photon event is then estimated by the Electronics Unit, 
and is necessarily corrected for in the calculation of the centroid.  
In order to mimic an integration-style data collection, we have included 
an imaging parameter to slow the scan rate of the CMOS by inserting a 
gap-time (i.e. a delay) between read scans of successive CMOS rows.  
The result is that when starting an imaging sequence, the first image 
will have a ramp-up to the total integration time by the end of the 
last row-read, but successive frames after that will all have equal 
integration times for each CMOS pixel row.  This is not as ideal as 
integrating the entire frame at once and then reading it all at once, 
but is quite workable.   

The maximum row-gap gives an integration time for the entire frame of
about 1.7 seconds (see Table 1). Generally, such integration cannot 
be done at high gain because the CMOS 
pixels will saturate almost instantly with most sources.  Therefore, a 
lower gain on the MCP is used in order to reduce the signal level
by a factor 10 or more, and 
therefore allows multiple photons events to fall on one location on
the CMOS without saturating the pixel wells.  True dark-current then becomes a problem.
Dark-current is typically highly dependent upon temperature, but the 
expected temperature averages of the detector in orbit are expected to 
be similar to those in the laboratory, typically around 20$^o$C.  These
measurements were therefore taken in order to examine some typical 
dark-current characteristics.  Table 1 shows the dark current from the three channels for different integrate times. The current is not uniform across
the detectors, so we show the range and mean, smoothed over 10 pixels. 
Table 2 shows the photon-counting read rates for different window sizes.

\section{Flat fields}

Flat fields were measured using a UV-reflective integrating sphere.  
The exit aperture of the sphere was 3 inches (significantly larger 
than the 40mm detector windows), and the detector window 
was placed directly in front of the aperture at a distance of 1-2 mm.  
Flat fields were measured for 5x5 Square and 3x3 Square centroiding 
algorithms, in photon counting mode, and at high and low voltage for each.  
Aside from photon statistics, at a particular voltage there would not 
be any expected broad-scale (i.e. larger than 1 pixel) difference between 5x5 
or 3x3 algorithms; the important differences should be seen at sub-pixel
resolution where fixed-pattern-noise effects are present. There is an 
expected broad-scale difference between high and low MCP gain (voltage) 
however, due to areal variations in MCP gain, as can be seen in Fig 1.  
At low gain, individual photon events on the CMOS are smaller than a 
readout pixel, so that the
centroiding fixed pattern is extreme. We describe this, and its correction,
in the following section. The large-scale flat field images are very 
uniform for nominal gain, but are significantly peaked in the centre 
at low gain.

\section{Fixed pattern noise}

There are several factors that affect the accuracy of the sub-pixel part 
of the centroid, as calculated.  The first is because the sampling 
gradation of the event is so coarse, i.e., it is significantly 
under-sampled, and this automatically causes the centroid to 
become systematically less accurate as the event centre lies closer 
to a pixel boundary.  A centroiding `box' of one 
pixel allows no sub-pixel centroiding; a box of three pixels in one 
dimension improves upon that significantly, but is still relatively
under-sampled, and a sample of 5 only improves slightly because very 
little energy is deposited in the outer pixels, and noise of various 
types becomes significant.

A second reason is that a linear centroiding scheme is being used to 
calculate the centre of a generally Gaussian signal.  This also causes 
a systematic error in the calculation of the centroid.  Gaussian 
centroiding schemes were tested in the initial phase of the 
Electronics Unit design, and were found to be extremely accurate at 
centroiding the expected event profiles.  However, these schemes 
require logarithmic calculations and were too slow for the FPGA hardware 
which was limited in how many cycles were available for each centroid
calculation, in order to keep up with the required high data rates.

The third, but possibly not final reason, is that an individual CMOS pixel 
is not uniformly sensitive across its area - some small areas of a pixel 
are more sensitive than other areas, and this is also a general systematic
effect.

Also, if the source of photons is non-isotropic, meaning that the 
distribution of photons on the detector face is not "flat", such as an 
extended source with flux gradients, or a point source convolved with 
the relevant resolution function, this will also introduce topological 
effects in the centroid histogram.  However, these effects are due to 
the source, and generally this is the scientific information one desires, 
and it is also not systemic.  The goal of course is to be able to correct 
for the systemic effects, while leaving the source-topological effects in 
place.  For this reason a flat field source, as provided by the integrating
sphere and the resulting images, is used to determine the systemic 
centroiding effects.

These systematic centroid errors are called Fixed Pattern Noise, or FPN.  
It would be somewhat more accurate to call it fixed pattern bias.
Figure 1 displays the effects of the Fixed Pattern Noise.  
At nominal MCP gain, the 3x3 algorithm
produces a distinct "screen-door" appearance, even though the source was 
illuminating the detector face evenly.  For the 5x5 algorithm the effect 
is somewhat less, but still apparent.  If the average FWHM of the events 
were only 0.1 pixels or more wider, the 5x5 algorithm performs much better 
than the 3x3, as shown in paper 1. However, the smaller events sizes
result from the deliberate choice of small window gaps, to yield the best
spatial resolution. At low MCP gain, both algorithms perform equally poorly. 

The resolution of the photocathode (see following sections), and the
resolution of the UVIT telescope are of order 0.4 readout pixels, so 
the flight science images will be centroided to 0.25 pixels. If we
centroid the flat fields to 0.06 pixels or better, we can resolve the
MCP pores, as shown in Fig 2. This is of use in that the pore grid can
be used as a measure of the centroid error and the correction
algorithms.

\section{Centroid error corrections}

The systematic centroiding bias, or Fixed Pattern Noise, can be corrected 
for significantly, while leaving the scientific information intact.  
This is simplest to do for the 3x3 Square centroiding algorithm because 
a mathematical solution can be programmed which closely mimics the 
real-world behaviour of the algorithm.  In the absence of noise, the 
3x3 Square algorithm would always return a centroid with a fractional part 
which does not fall outside of the maximum-pixel boundary and can be 
thus corrected.  For the 3x3 Square, it is rare for noise effects 
to skew the centroid outside the boundaries of the maximum pixel, and 
therefore the noiseless mathematical solution is a decent approximation 
of the required correction. 

The situation is less good for the 5x5 algorithm, since noise effects 
in the outer pixels with low signal, skew more
centroids outside of the central maximum pixel, and 
therefore a Monte-Carlo style simulation would be required in order 
to develop a correction algorithm numerically. On the other hand, 
as we showed
in paper 1, the 5x5 algorithm gives better controids without correction.

The mathematical solution is simply the comparison of a theoretical 
probability distribution of centroids, to the real-world probability
distribution.  If the centroiding accuracy were perfect, then the 
probability distribution of intra-pixel centroids would simply be a 
straight line of slope equal to the ratio of the total number of events 
over the number of sampling bins within a pixel.  The corrections need 
to be performed at a much higher resolution than the physical sampling 
(1/16th of a pixel), and so the real centroids are redistributed evenly 
within their 1/16th bins, at 1/1024th pixel resolution.  The cumulative 
sum of this new distribution is its cumulative probability distribution, 
and this can be divided by the constant-slope line of the theoretical
probability distribution to get the correction factor for each 1/1024th 
bin.  In other words, each 1/1024th bin of the interpolated centroid 
data is multiplied by the correction factor, which is the ratio of the 
real cumulative probability distribution over the theoretical one.  
If the real data were composed of centroids from a "flat" illumination of 
photons, then the systematic centroiding bias effects will have been 
corrected for.  This flat-field correction curve can then be applied 
to data which has sources and other topological variations - i.e. 
valuable scientific information, and the correction will only apply 
to the existing systematic bias within the data, but leave the
source-topological variations undisturbed.

Paper 1 gives more detailed description of the corrections, and shows
how the signal is redistributed within a pixel. Essentially, the distribution
is remapped by performing a division by the correction factor. All photons 
are conserved, so no photometric error is introduced. 

The fraction of out-of-pixel centroids depends to some extent on the
size of the photon event on the CMOS readout. For our flight detectors,
with the 3x3 Square algorithm, there are approximately 1-2\% of centroids 
which fall outside of the central pixel boundary, and therefore these 
will not be adjusted to the correct position and will thus constitute 
noise.  For the 5x5 Square algorithm, the fraction is 
approximately 5-7\%.  However, it should be pointed out that the reason 
these centroids were outside of the pixel boundary in the first place 
was because of a noise skew, and so the correction scheme doesn't 
actually introduce more noise, but rather simply conserves it.

Figures 2 and 3 show parts of flat-field images with 3x3 centroiding
before and after correction. The final science images will have the 
1/4 pixel resolution of Figure 2. A final point to note is that in 
flight, there will be spacecraft jitter of amplitude a pixel or more
over the time of an observation. This will be monitored by guide-star
centroids from the VIS channel and the photon centroids corrected 
in building up the science image. The effect of this will be to
smooth out the FPN, so that it may be a matter of post-observation
observer choice whether the corrections are needed in a given image.

\section{Hole mask data}

Spatial resolution and flux calibration were performed using the hole mask 
(see Fig 4). These holes are all 25 microns in diameter, to an accuracy 
of 5\%. The mask was taped close to the detector face, about 1mm from
the window, and illuminated 
with the collimated beam. There are two points to note in this. First, the 
beam illumination underfills the detector window, and is not uniform. 
Second, the mask is a small distance from the photocathode, by the window
thickness and the gap between the window flange and its outer surface, so 
the spots are somewhat enlarged by diffraction. 

We used the first point to our advantage by the fact that the spot images 
cover a wide range of signal levels, thus increasing the dynamic range of 
the flux calibration. The response uniformity across the detector window 
was checked by using a range of flux levels, and by moving the detector 
around the collimated beam.

We calibrated the second point by using results from a different hole mask 
used in testing the engineering model detector. This mask had alternating 
hole sizes of 25, 50, and 75 microns across its diameter. The actual spot 
image sizes are shown in Fig 5. Given the very linear relationship,
it is evident that the diffraction and spot size effectively add, and
we have used the extrapolation to zero hole size as a measure of the
detector resolution, as shown in the diagram. The measured spot sizes
for the 25 micron hole mask data were similar to that in Fig 5. The same 
slope and extrapolation was therefore applied to the image size measured 
with the 25 micron hole-mask used with the flight detectors. There was
insufficient calibration time to get images with the multi-sized hole-mask 
with the flight detectors, but the geometry and window thickness was 
the same for all detectors, so Fig 5 was applied to all cases. In the next
section, we note other ways to make this measure, and their good agreement.

Spot sizes are defined as the FWHM values from the images. These are 
expressed in readout pixels, and can be converted to values at the 
photocathode or arcseconds on the sky as seen with the UVIT telescope. 
Signal levels were integrated over the final spot images. 

\section{Spatial resolution}
 
The spatial resolution was also estimated by the images of the cross-shaped  
hole arrays with differently spaced separations. 
These arrays were oriented at several different angles, to allow different
sampling by the readout pixel array.  Figure 6 shows some of these images, 
not corrected for centroiding systematics, and Fig 7 shows plots from 
summing pixel rows across one of the cross arms.

The smallest spot separation is 30 microns, centre to centre, so the 
smallest gap between them is 5 microns. The next smallest gap is 20 
microns. The resolution is different for the 3 detectors, as expected 
from their window gaps and the photon wavelengths (see figs 7 and 8), 
but in all 
cases the 20 micron gap is well resolved, and the 5 micron gap is not. 

The resolution estimates by both methods agreed well, and table 3 
shows the adopted average numbers for the spatial resolution for the 
3 detectors.  The wavelengths used, and the photocathode gaps are also 
shown. The `predict' numbers are the values expected from the front-end tube
measures. Figure 8 shows these and the dependence of resolution on 
wavelength and gap, and compares 
the front-end tube tests before assembly with the final values from the 
fully integrated detectors.  The numbers `corrected to sky' give our estimates
of FWHM in arcseconds on the sky. The values are much as expected, except for 
the visible wavelength detector, which is about twice the expected value.
The longer wavelengths of the VIS channel will cause more diffraction 
through the mask apertures but will not explain this amount of difference. 
The scientific impact is small, however, as the telescope optical 
resolution is poor because of passing through a beamsplitter. The main 
use of the optical channel is for tracking spacecraft drift via star 
centroids, and there will be no impact on this performance, since the 
much larger integrate mode images will be used.

We note also that the resolution as measured from images with low MCP gain 
is very similar. The percentage numbers given for these are the signal 
reduction when using low gain, as it is intended for use with bright stars 
which would cause detector damage at full gain.

In Figure 9, we illustrate the resolution in the case of two closely spaced
point source images. The profile is approximated by a gaussian, but an average
spot image is shown, and has a slightly narrower peak and some weak broad wings.
The plot shows how the profile is signicantly broadened with a separation of
only some 0.3", and shows separate peaks when separated by about 0.8".

These results indicate that the UV detectors will yield image resolutions 
of 1 arcsec or better, and the visible channel a little larger. The 
ultimate resolution of the UVIT instrument will combine these with 
the optical resolution of the telescope and the drift of the spacecraft 
during observations. The telescope should deliver images of 1 arcsec FWHM, 
and we expect to correct drift using the visible channel guide-star centroids, to a small fraction of that.

Figure 10 compares the integrate mode images (simply the sum of all the 
photon event splashes) with the centroided image resolution we report 
above. Integrate mode images can be centroided to 0.1 pixels, or 0.25 
arcsec, to compensate for spacecraft drift.

\section{Flux calibration}
 
The flux calibrations used the same hole-mask data. The wavelength was 
selected with the monochromator, and the flux monitored by photometer 
readings from a beam-splitter. A series of (usually four) flux settings 
was used for all images. Data were taken in photon-counting mode with 
5S and 3S centroiding, and Integrate mode, and with full frame (29Hz) 
and a subwindow (293Hz) readout.  Images were accumulated for all these 
settings, and their relative count rates logged by measuring the event 
counts per second (not per frame) of the same (brightest) spot. The 
sub-arrays were chosen to cover a region well-filled with measurable 
spots. In any images, there are spots covering a wide range (about a 
factor 5) of signal levels. The spots are small and well-separated, so 
no crowded field issues arise, and they simulate star images in a 
sparse field. Having established that there is no wavelength dependence 
of signal saturation, one wavelength was selected for this work for each
channel. In addition, all data were taken at nominal and low gain. 

In any one image, about 6-10 spots were measured, covering the full 
range of signals. The same spots were measured for all four flux 
settings. The signals were measures using the `rimexam' task within IRAF, 
using radii selected for the spots for object and surrounding background. 
The same measurement settings were used for all spots and channels. The measurement radius was 5 pixels, and 
the sky an annulus between radii 7 and 9 pixels. The signal levels were
accumulated in a table for all spots and flux settings.  

The saturation plots (Fig 11) were generated as follows. Signal 
level is plotted against photometer flux reading for the brightest spot measured. The flux levels of the fainter spots were scaled by the ratio 
of the signal levels for the lowest two flux settings, compared with that 
for the brightest spot. In all cases the signals were not saturated for 
the lowest two fluxes, so this scaling should be good. This leads to 
points that cover the range of flux and signal quite well. In the case of 
the subarrays, the number of spots was more limited, so that it was 
sometimes not possible to cover the ranges as well as we wished.  

Figure 12 shows an EM NUV channel saturation plot, derived from the
mask with different-sized holes. The saturation curve is different
from those of the flight systems 
because of the different event size and poorer resolution of this
detector, illustrating how individual each system is. Figure 12 also 
shows the flux calibration for VIS channel integrate mode images, 
whose effective integration time is set by the time gap between 
pixel row reads (row gap). This low gain INT mode operation is 
expected to be the default for the VIS channel when used for drift
monitoring on bright guide stars. 

There are similar results for the NUV spot signals with low gain setting. 
For all low gain operations, we note that the flat fields show large 
variations in signal from centre to edge of the field, so that flux 
calibrations will be reliable only for objects in the central part. 
In addition, low gain is intended for cases where there are bright 
objects giving high count rates. We can adopt low event thresholds to 
account for the lower gain, but this allows noise to be included in the 
accepted data. The result of these considerations is that the calibration 
plots, shown below, show significant scatter and non-linear response at 
low signal levels.  Nevertheless, they are useful in allowing observations 
of bright targets, which should be placed at the centre of the field and read
with a small subarray. 

So far, we have used only signal levels in counts from the images as 
described, and flux levels as measured in arbitrary units on our
photometer, via a beamsplitter. In order to convert signal levels to 
photons at the photocathode, we need to apply a factor that measures 
the absolute QE of the systems. This was done by using a NIST calibrated 
diode in the beam, and the results are shown in Figure 13. 
This way we can convert from source counts to
incident photons. To plan observations through the entire telecope
we need to include an extra factor for the telescope optics and
filter throughputs as well. These will be published and updated 
on the proposal website for the mission, once established. 

The absolute QE of the FUV detector at the wavelength of observation (150nm) 
is measured at close to 3\%. This gives a conversion from the source
total signal in the plots to photons at the photocathode of 0.77. That is,
we multiply the source counts by 0.77 to get the incident photon counts. 
The NUV wavelength used was 210nm, at which the QE is close to 10\%, and
the QE for the visible is also close to 10\%. This gives a conversion 
factor from plotted source signal to photons of 0.37 for these channels. 

\section{Detection of double-photon events}

As noted in paper 1, double-photon events can be flagged by using the 
max-min corner pixel values from the 5 x 5 events boxes. It is these 
events that lead to the non-linearity and saturation plots shown.  
We performed some double-photon rejections based on various values of 
the rejection threshold, to check whether saturation leads to degradation
of the spatial resolution.

The reader is referred to paper 1 for the detailed description of this
rejection process, but we give a brief summary here. Double-photon events
statistically lead to broader events patterns in the readout image, and
they can be isolated by the difference between the maximum amd minimum
pixels signals in the 5x5 event box. Each detector has an optimum
value of this threshold, depending on the individual event sizes and 
the intrinsic spatial resolution of the device. Using `no rejection'
accepts all events, while a high rejection threshold will identify only 
the most egregious of double-events. 

In the case of single isolated spots (i.e. stars in real observations), 
there is no significant change to the point source resolution if we 
reject such events. Thus, we do not need to apply such thresholds for 
science images to resolve stars. This is true for images of two closely 
spaced spots (stars). 

Figure 7 of the paper by Srivastava, Prabhudesai, and Tandon (2009), 
points out how 
flux can be lost from bright crowded regions of extended images with 
structure. To look for this, we inspected the close-spaced cross of spots 
in our images with a range of count rates. This will show up as spots 
in the centre of the pattern having less signal than the outer 
(less crowded) ones, since the spots imaged all in fact have the same 
flux. In general, we don't see any effect for the lower three flux values,
but it does show up in the highly saturated images. 

This is potentially a problem for science analysis, but can be alleviated 
by identifying the double events in the image, and adding them back in again, 
so they get counted twice, as they should be, instead of only once. In
principle, triple events can be treated in the same way by setting a 
different (lower) max-min threshold, but this is less reliable as it is 
more susceptible to random noise too.  Figure 14 shows an example of this
correction in one close array of spots, where saturation is clearly present 
in the crowded inner parts of the image.  Table 4 shows the flux 
corrections at two different flux levels, for isolated spots, the
end spots of the cross-grid, and the entire cross-grid. The values 
are shown for a range of rejection threshold values. Isolated spots
show no signal loss at the low flux value, while at high flux, they do - more markedly for the brighter spot, as would be expected. The cross-grids
behave in many ways like an extended object with structure, and signal loss
(double events) are seen even in the outer spots. The signal loss
of the entire grid is seen at both flux values. The table also
gives the restored flux from compensating for the double events by 
counting them twice. It is clear that this process will be needed
to analyze bright extended images. 

We thank Routes Asto-engineering, particularly Don Asquin and Josee 
Cayer for their work on the hardware and its initial testing. We also 
thank Shyam Tandon for his active participation, and discussion of the 
results presented here. We thank the referee for useful comments
on our presentation. We thank the Canadian Space Agency for support.

\newpage

\centerline{\bf References}

Breeveld A. A., et al, 2010, MNRAS, 406, 1687

Hutchings, J.B., Postma J., Asquin D., Leahy D., 2007, PASP, 119, 1152

Kuin N.P.M., and Rosen S.R., 2008, MNRAS, 383, 383

McPhate J., et al, 2010, Proc SPIE 7732, 2H

Morrissey P., et al 2009, ApJS, 173, 682

Morrissey P., et al 2005, ApJ, 619, L7

Srivastava M.K., Prabhudesai S.M., Tandon S.N., 2009, PASP, 121, 621

Vallerga J, et al 2002, Nucl Instrum Meth A 477, 551

\vskip 2cm

\newpage

\centerline{\bf Captions to figures}

1) FUV detector flat field images with 3x3 centroiding at 1/8 pixel
resolution: 8x8 pixel subarray detail at the top and full images on 
the bottom. The left side images are with nominal gain and the right 
side are at low gain.

2) NUV flat field detail with 1/32 pixel resolution, in which the MCP pores
can be seen. The centroids are from 3x3 subarrays. The right panel shows
the raw image, which shows the pixel boundary effects clearly. The left
image has been corrected and removes most of the errors. 

3) NUV channel 3x3 centroid detail at 1/8 pixel resolution. The left
panel is the raw image and the right panel has centroid corrections applied.

4) Details of hole mask used for flight system flux and resolution 
measurements. The holes are all 25 microns in diameter. The cross-shaped 
arrays have different orientations to sample the readout pixel array
differently.

5) Measured EM spot image FWHM from a hole mask with 25, 50 and 75 micron holes.
The numbers in parentheses are the actual image FWHM at the photocathode,
broadened by diffraction betwen the mask and the photocathode. The system
spatial resolution is defined as the size of the detected image from the
linear fit extrapolated to zero size. See text for discussion. 

6) Some typical 3x3-centroid images of the spot array in different 
orientations. The readout pixel edges are clearly seen. 

7) Plots of signal from rows summed along the arms of the spot array, with
detectors at wavelengths noted. The smallest separation is 30 microns.
The prolfies are slightly confused near the centre where they cross, and
by pixel boundaries in the individual spot images. The 180nm image
is the dotted line, and has the highest resolution (see Table 3).

8) The measured image FWHM in microns at the photocathodes. The angular 
scale on the sky is shown on the right. The lines are derived from 
`front-end' only resolution measures made on prototype tubes. The tube
resolution depends on the wavelength (photoelectron energy), and the 
gap between the photocathode and the MCP surface. The filled
squares show the measured flight system values, and the open square is 
the NUV engineering model. The final system resolutions are close to 
expectation except for the VIS channel.

9) Plots showing the profile from two calculated images represented by 
Gaussians of FWHM 0.8", shown as the dark solid line. The actual image 
profile for the FUV detector is the dotted line, which is slightly 
narrower at the peak, but with some outer wings. 
Thus, the model is a good representation. Significant profile broadening 
is seen at 0.3" separation and the peaks are separated at 0.8" separation. 

10) Comparison of the same spots imaged in photon-counting mode (left) and
in integrate mode (right). The photon-counting images are about 1" on the sky 
and the integrate mode images are many times broader (FWHM $\sim$10" 
on the sky).
However, the integrate mode images can be centroided to 0.2" or better
to correct spacecraft drift at that level. in building up the photon-counting
images.

11) Flux calibration of FUV detector, comparing saturation effects
between slow and fast read, and normal and low gain operation. In the top
two panels, dots are from high gain and circles from low gain. Dotted lines
indicate the non-linear saturation effects. Rapid read subarrays allow
a alrger dynamic range before saturation becomes severe. Low gain reduces the
signal levels by factors two or more. The text in section 8 describes the
units and their conversion to photons. 

12) Upper: saturation for the EM detector. Signals are in units of
1000 counts. The different symbols refer
to small and large hole spots on the same mask, scaled by their relative
throughput. Lower: Flux calibration for the VIS channel detector with
integration time (row gap), for two different flux levels. Both are in the
low MCP gain mode. 

13) QE of the three detectors with wavelength. 

14) 3-d representation of spot array image at high flux level. 
The spots all have the same brightness. Double events
in the crowded central area reduce the inner spot signals (top). 
The middle panel shows how many single events are present. Tthe
lower panel shows the restores flux image from adding back the double 
events to the image. The lines across the peaks are to guide the eye.  

\newpage

\begin{deluxetable}{cllllllllll}
\tablecaption{Dark current values in electrons per read}
\tablehead{\colhead{Row} &\colhead{Read}\\
\colhead{gap} &\colhead{interval} &\multicolumn{3}{c}{FUV} 
&\multicolumn{3}{c}{NUV} &\multicolumn{3}{c}{VIS}\\
\colhead{0.0033sec} &\colhead{Sec} &\colhead{Min} &\colhead{Max} &\colhead{Mean}
&\colhead{Min} &\colhead{Max} &\colhead{Mean} &\colhead{Min} &\colhead{Max} &\colhead{Mean}   }
\startdata
0 &0.035 &1339 &1595 &1390 &1442 &1727 &1497 &1447 &1733 &1512\\
63 &0.241 &1305 &2577 &1362 &1398 &2641 &1451 &1420 &2826 &1488\\
127 &0.451 &1308 &3438 &1373 &1375 &3441 &1444 &1423 &3788 &1504\\
191 &0.661 &1307 &4133 &1395 &1368 &4122 &1455 &1423 &4569 &1528\\
255 &0.870 &1307 &4705 &1424 &1366 &4723 &1477 &1427 &5244 &1561\\
319 &1.080 &1311 &5198 &1460 &1371 &5232 &1505 &1432 &5817 &1598\\
383 &1.290 &1318 &5624 &1501 &1377 &5636 &1542 &1437 &6315 &1640\\
447 &1.500 &1322 &5988 &1546 &1385 &5913 &1582 &1446 &6757 &1685\\
511 &1.709 &1330 &6289 &1594 &1392 &6253 &1625 &1456 &7150 &1733\\
\enddata
\end{deluxetable} 

\begin{deluxetable}{lll}
\tablewidth{5cm}
\tablecaption{Read rates for different subwindow sizes}
\tablehead{\colhead{Width} &\colhead{Height} &\colhead{Hz}  }
\startdata
512 &512 &28.7\\
409 &409 &44.3\\
306 &306 &77.3\\
203 &203 &168\\
100 &100 &605\\
512 &100 &147\\
100 &512 &118\\
\enddata
\end{deluxetable}

\begin{deluxetable}{lllllll}
\tablecaption{Spatial resolution summary}
\tablehead{\colhead{Property} &\multicolumn{3}{c}{FUV} &\colhead{NUV}
&\colhead{VIS} &\colhead{EM}\\
&\colhead{180nm} &\colhead{150nm} &\colhead{125nm} &\colhead{210nm}
&\colhead{400nm} &\colhead{230nm} }
\startdata
Gap {($\mu$)} &110 &110 &110 &100 &140 &170\\
Event size (px) &0.92 &0.92 &0.92 &1.18 &1.16 &1.30\\
Image FWHM (px) &0.35 &&0.45 &0.38 &0.60 &0.55\\
Expected\tablenotemark{1} &0.8" &&1.1" &0.7" &0.7" &1.1"\\
FWHM on sky &0.8" &&1.0" &0.9" &1.3" &1.2"\\
\\
\bf{Low gain}\\
Signal level\tablenotemark{2} &&10.5\% &&8.6\% &4.8\%\\
FWHM (px) &&0.34 &&0.35 &0.53\\
\enddata
\tablenotetext{1}{From the lines in Figure 8}
\tablenotetext{2}{Signal level reduction from nominal gain}
\end{deluxetable}

\begin{deluxetable}{rrlrr}
\tablecaption{Source counts in images with double-events\tablenotemark{1}}
\tablehead{\colhead{Rejection} &\colhead{Spot1} &\colhead{Spot2}
&\multicolumn{2}{c}{Cross-grid}\\
\colhead{threshold} &&&\colhead{End spot} &\colhead{Whole pattern} }
\startdata
&\bf{Medium flux =} &\bf{0.66 events/read}\\
None &794 &1476 &534 &11161\\
128 &797 &1459 &536 &7781\\
64 &811 &1446 &450 &7126\\
16 &766 &1343 &269 &4562\\
Restored\tablenotemark{2} &&&608 &12785\\
\\
&\bf{High flux =} &\bf{1.04 events/read}\\
None &2838 &4996 &2106 &35820\\
128 &2739 &4590 &815 &6877\\
64 &2657 &4165 &512 &4330\\
16 &2369 &2724 &46 &1061\\
Restored\tablenotemark{2} &&&3419 &64765\\
\enddata
\tablenotetext{1}{See text for detailed explanation}
\tablenotetext{2}{Restored image signal by counting double events twice}
\end{deluxetable}
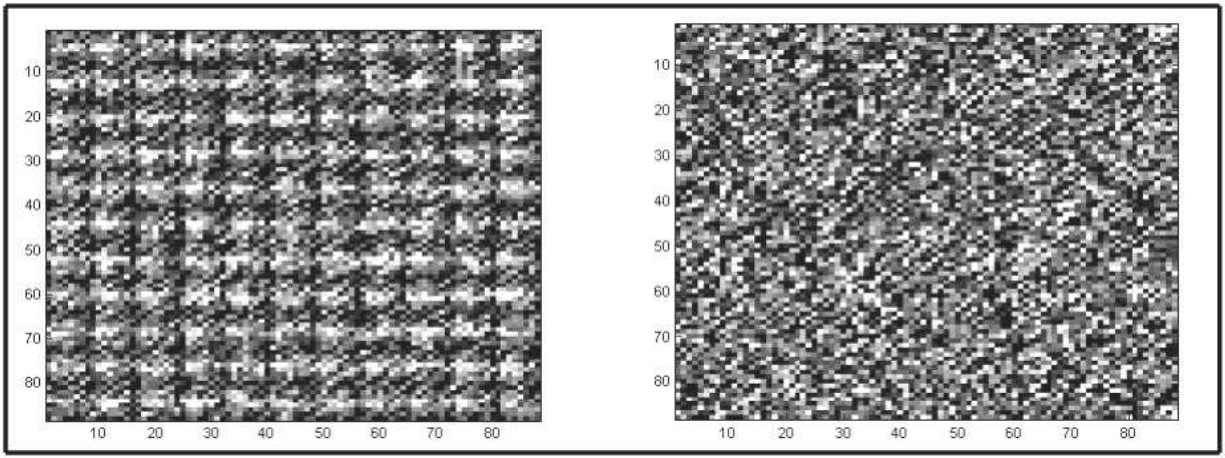
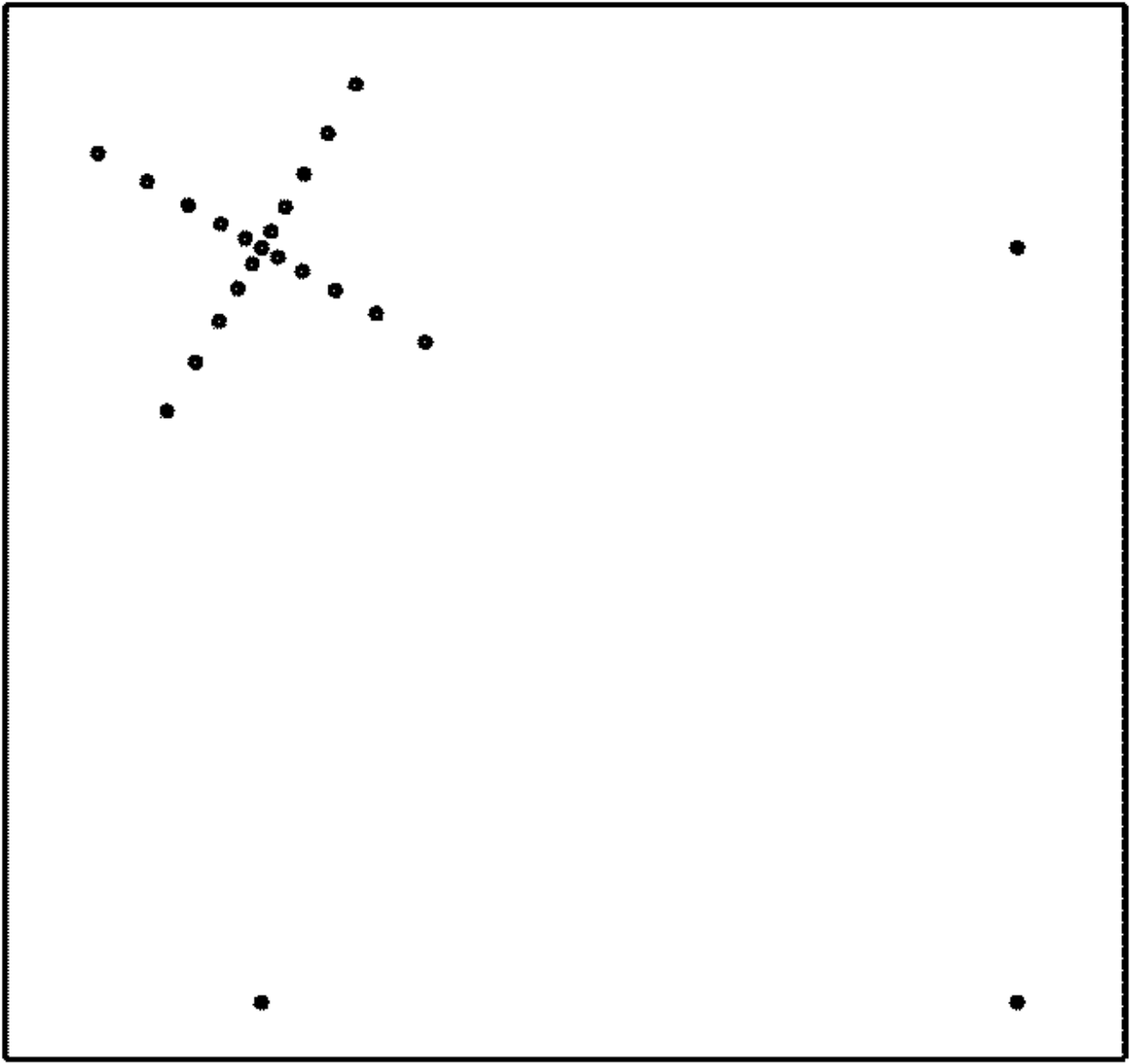
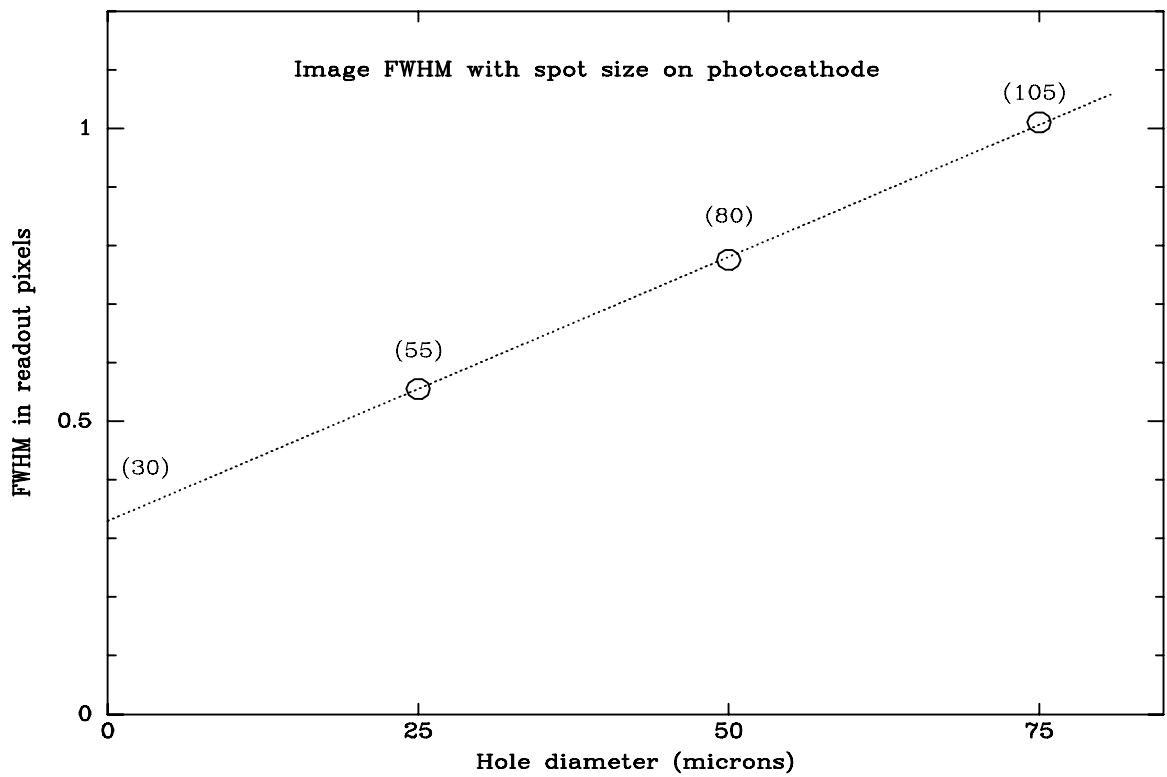
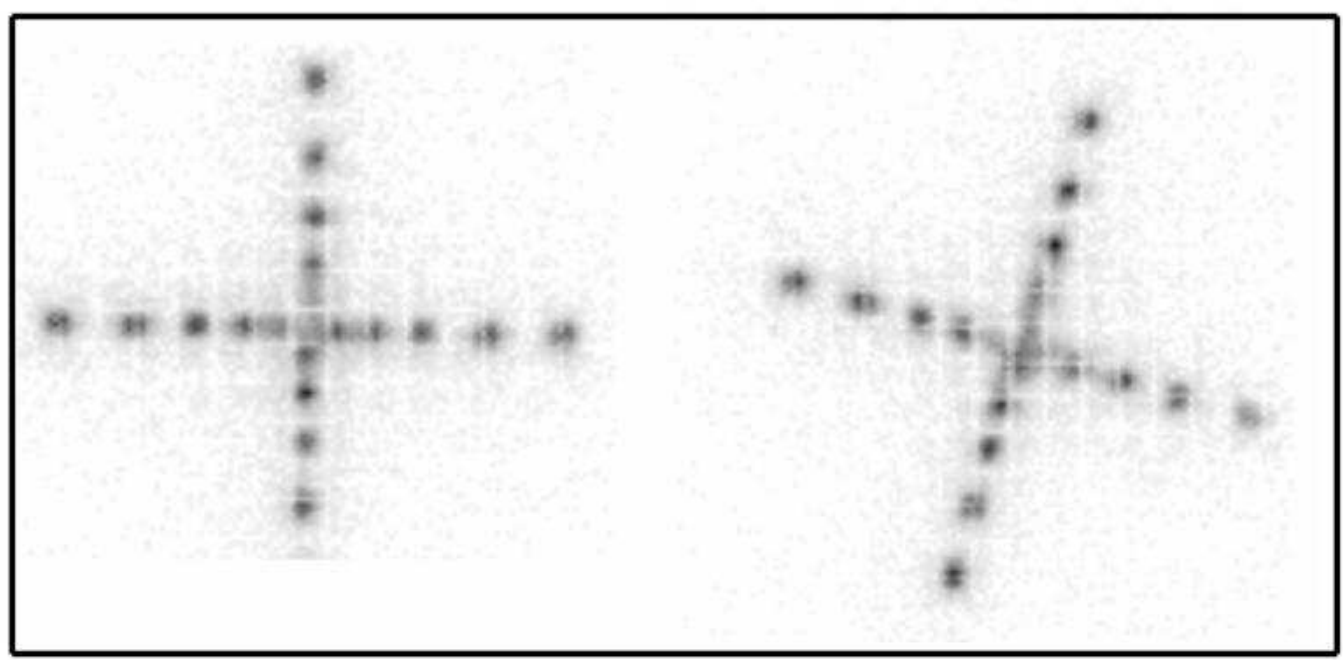
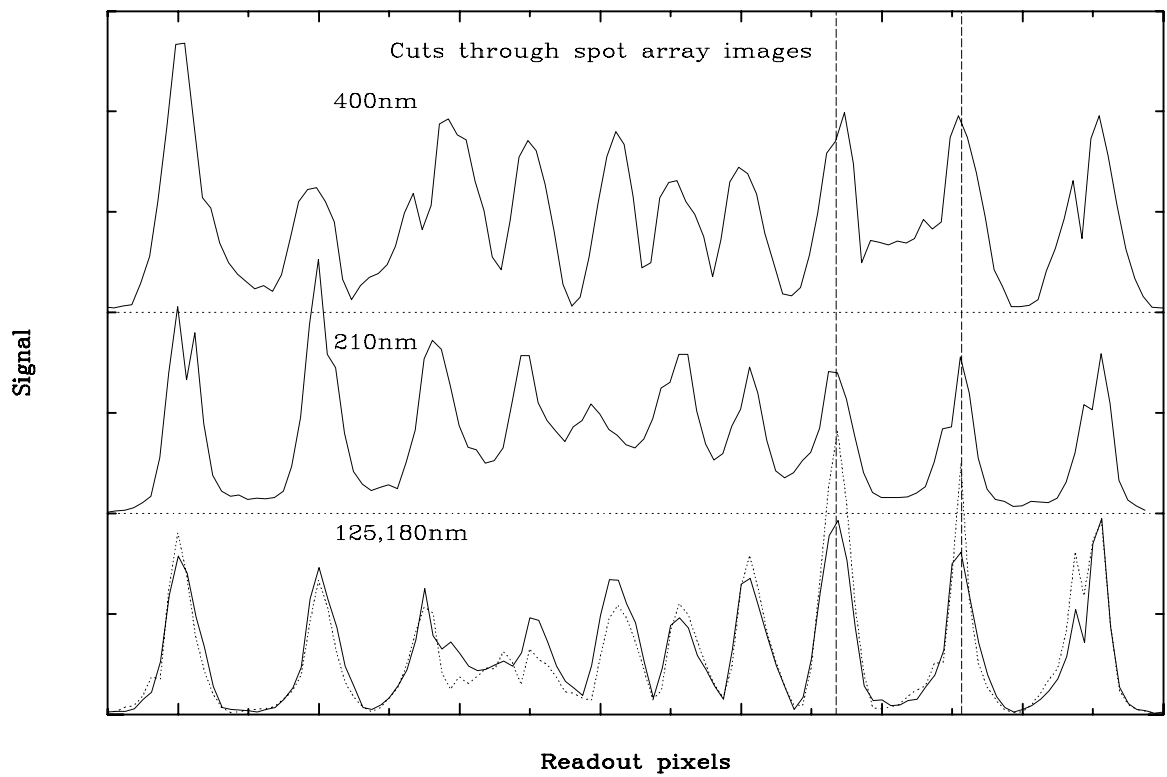
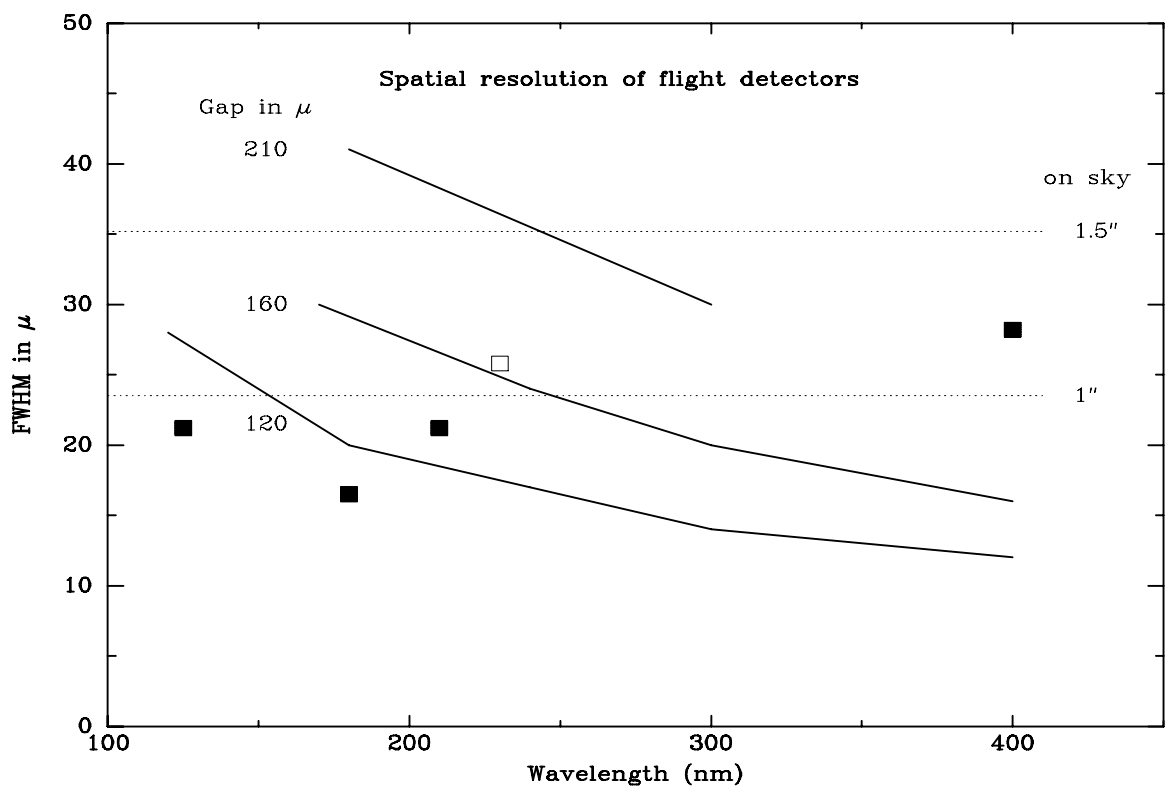
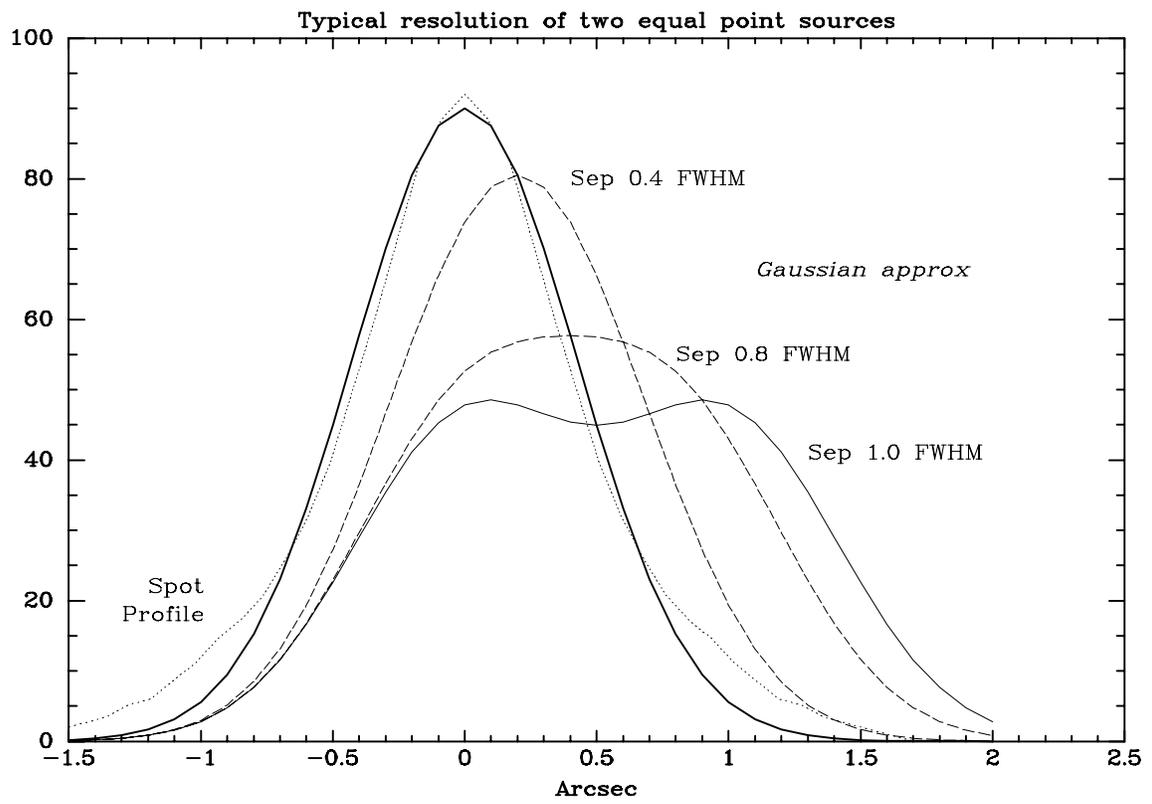
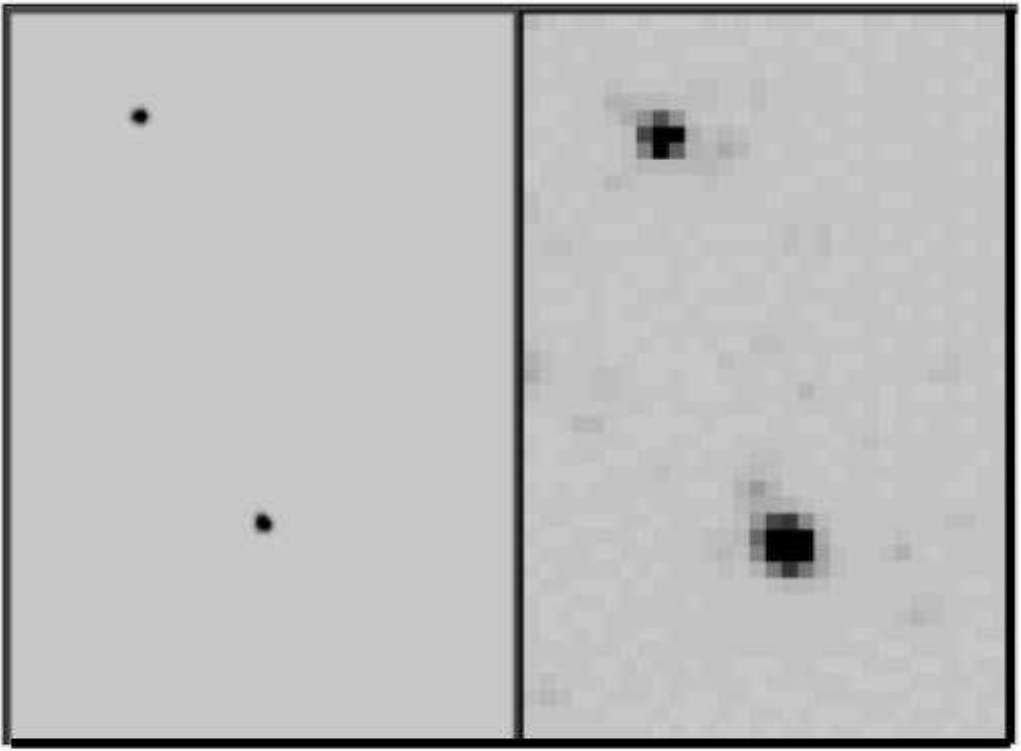
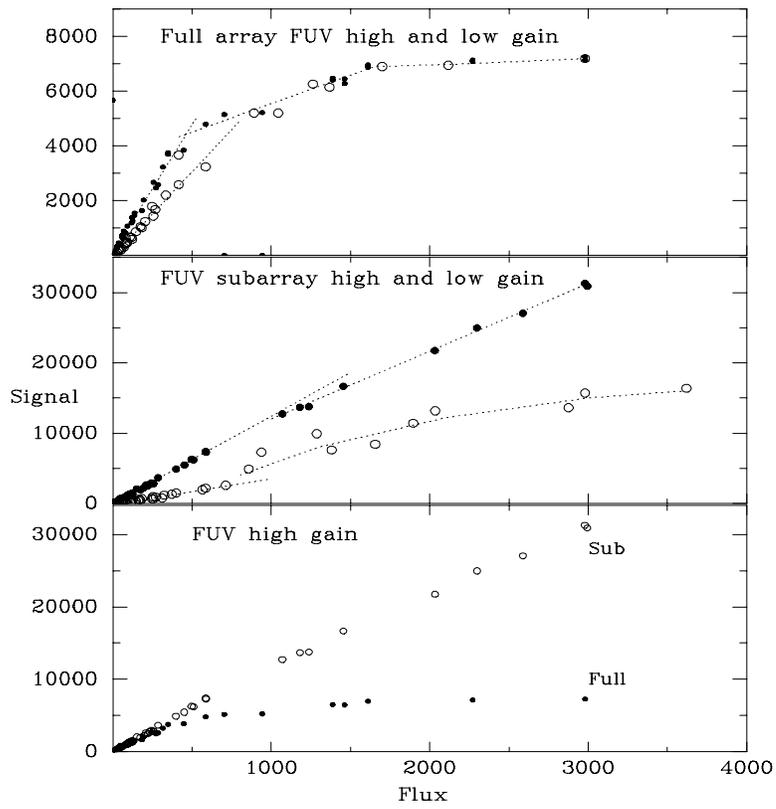
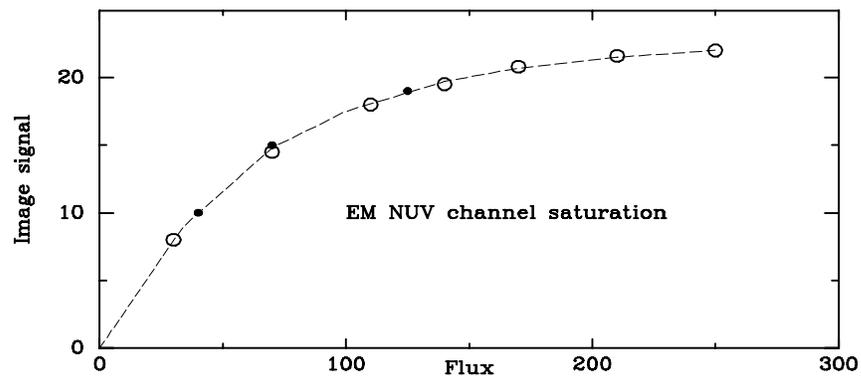
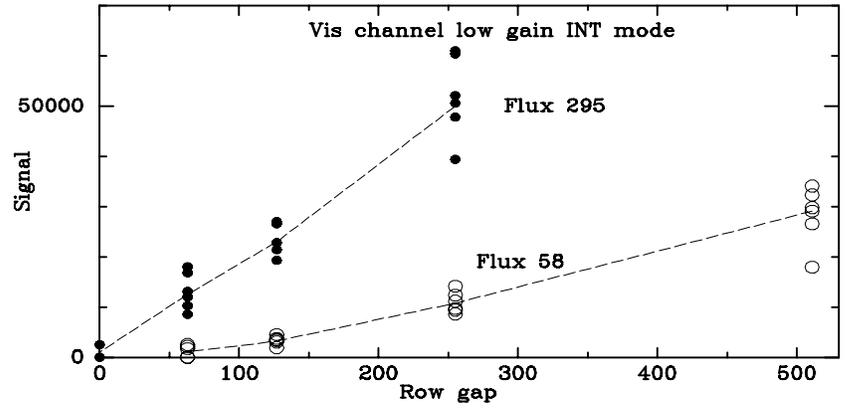
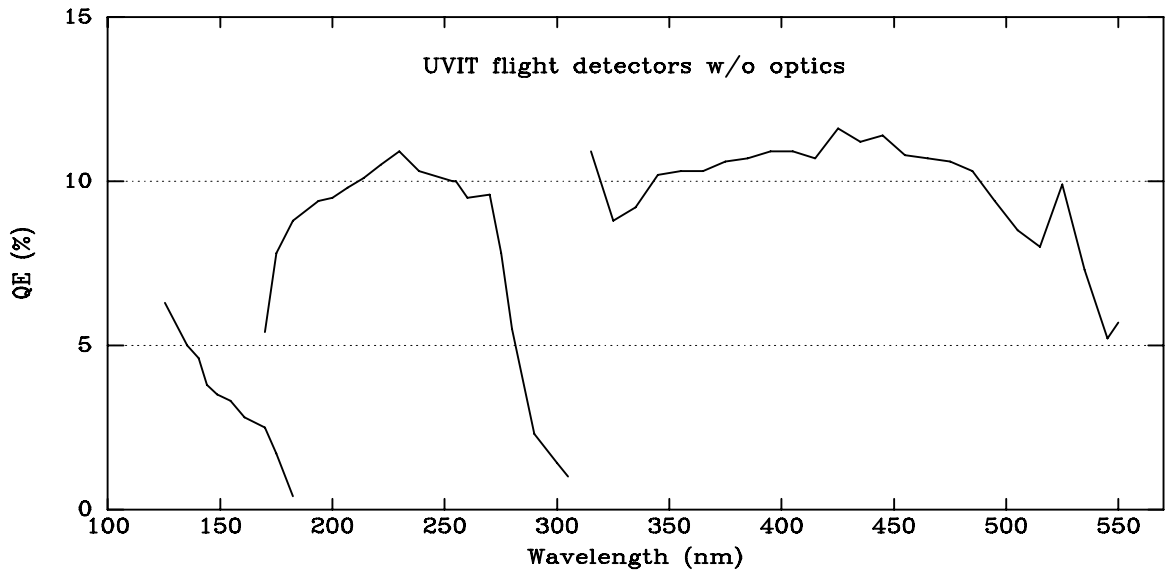
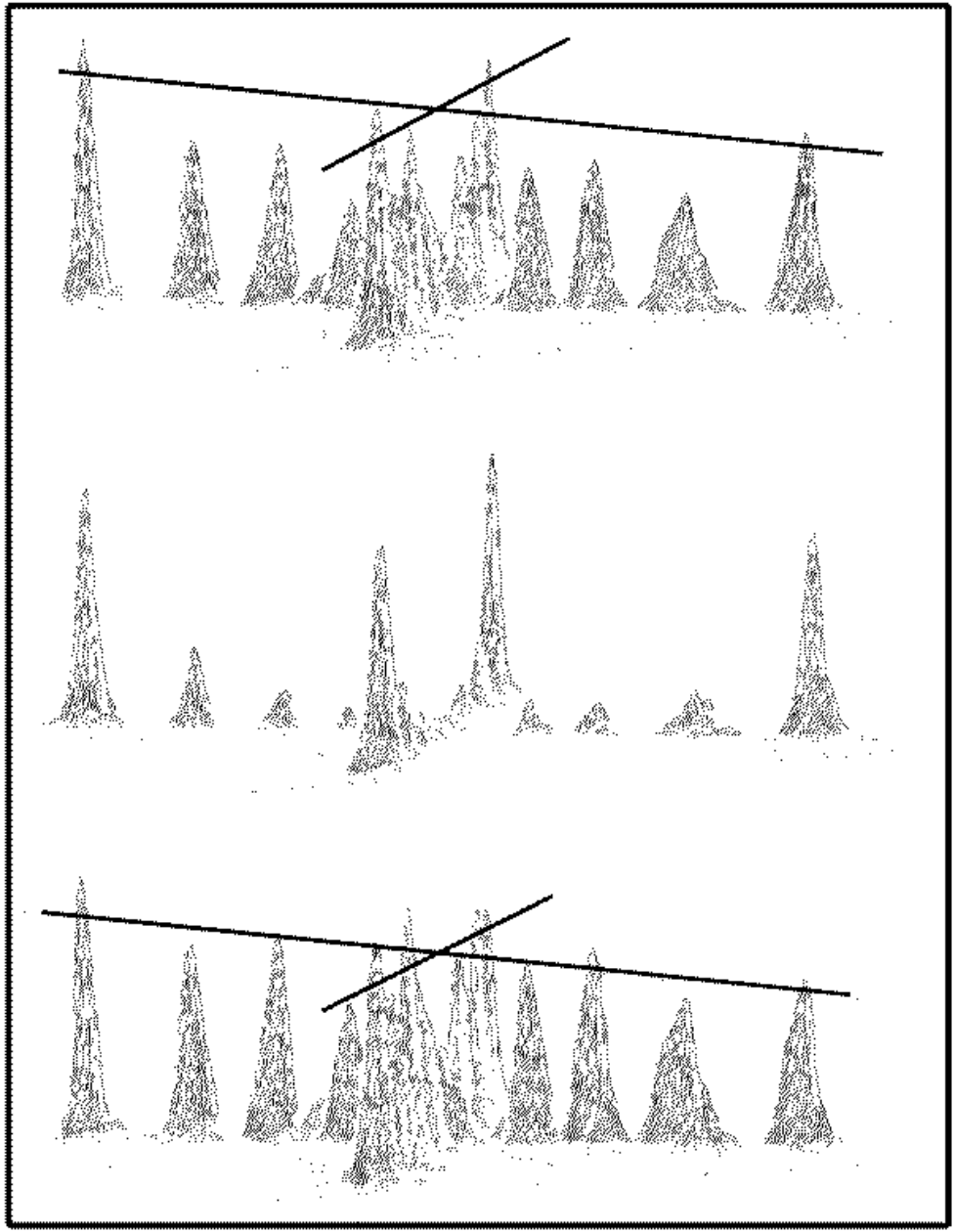
\end{document}